\documentclass[twocolumn]{autart}

\usepackage{graphicx}
\usepackage{graphics} 
\usepackage{epsfig} 
\usepackage{mathptmx} 
\usepackage{times} 
\usepackage{amsmath} 
\usepackage{amssymb}  

\def\sn{\mbox{$| \! | \! |$}}

\def\R{\mbox{\mathbb{R}}}

\def\rT{{\rm T}}        

\newcommand{\rank}{\mathop{\mathrm{rank}}\nolimits}

\newcommand{\diag}{\mathop{\mathrm{diag}}\nolimits}

\def\rank{\mbox{\rm rank\,}}

\def\R{\mathbb{R}}

\def\be#1{\begin{equation}\label{#1}}
\def\ee{\end{equation}}

\def\H2{\mathcal{H}_{2}}
\def\Hinf{\mathcal{H}_{\infty}}

\newtheorem{theorem}{\bf Theorem}
\newtheorem{definition}{\bf Definition}
\newtheorem{lemma}{\bf Lemma}

\newtheorem{remark}{\bf Remark}
\newtheorem{problem}{\bf Problem}

\def\rank{\mbox{\rm rank\,}}
\def\sym{\mbox{\rm sym\,}}

\begin{document}

\begin{frontmatter}
%
\title{Robust State-Feedback $\Hinf$ Control For Discrete-Time Descriptor Systems With Norm-Bounded Parametric Uncertainties}

\thanks[footnoteinfo]{The material in this paper was not presented at any IFAC meeting.
 Corresponding author is A.~A.~Belov, a.a.belov@inbox.ru}

\author[ITMO,ICS]{Alexey A. Belov}\ead{a.a.belov@inbox.ru},
\author[ICS]{Olga G. Andrianova}\ead{andrianovaog@gmail.com}

\address[ITMO]{ITMO University, 49 Kronverksky Pr., St. Petersburg 197101, Russia}
\address[ICS]{V. A. Trapeznikov Institute of Control Sciences of Russian Academy of Sciences,
65 Profsoyuznaya street, Moscow 117997, Russia}  

\begin{abstract}
This paper deals with a state feedback $\Hinf$ control problem for linear time-invariant discrete-time descriptor systems with norm-bounded parametric uncertainties. To this end, bounded real lemma (BRL) is extended on the class of uncertain descriptor systems. The control design procedure based on the conditions of BRL for uncertain descriptor systems is proposed. Numerical example is included to illustrate the effectiveness of the present result.
\end{abstract}

\end{frontmatter}

\section{Introduction}

Descriptor systems also referred to as singular systems become a subject of interest for many researchers in the last few decades. It turns out that descriptor systems representation is a general case of a state-space one. Descriptor systems found their various applications in different sciences, for example, in economics, biology, electrical engineering, etc (see \cite{Dai1989,Duan2010} and references therein). Mathematically a descriptor system is a combination of dynamic and algebraic equations. In discrete-time case, a specific behavior such as non-causality may occur while solving system's equations, due to presence of algebraic constraints in the system's model. Motivated by this fact, many efforts have been made towards developing methods to solve a number of control problems. Among them are admissibilization, filtering, control design, etc \cite{Dai1989,Xu}.

Problems of sensitivity reduction or external disturbance attenuation are well-known in modern control theory. The $\Hinf$ control problem is one of the most popular among them. In this case, a disturbance is assumed to be a square summable sequence. The $\Hinf$ norm of the closed-loop systems stands for the system's gain from the input to the controllable output and is required to be less than a given scalar $\gamma>0$. There are a lot of different approaches for solving this problem for discrete-time descriptor systems \cite{Chadli2012,Feng2013,Rehm2002,Xu,Yung2008_1,Yung2008}. To this end, several versions of BRL have been proposed in literature to solve $\Hinf$ performance analysis problem. All of the results mentioned above deal with uncertainty-free discrete-time descriptor systems.

Robust $\Hinf$ control problem for discrete-time descriptor systems with parametric uncertainties is investigated in \cite{Chadli2014,Coutinho2014,Ji2007,Xu}. The paper \cite{Coutinho2014} deals with linear discrete-time descriptor systems with polytopic-type parametric uncertainties. The obtained conditions are \textit{bilinear}, so the iterative procedure of control design with $\gamma$-value minimization is given. In \cite{Chadli2014,Ji2007,Xu} norm-bounded uncertainties are under consideration. The result obtained in \cite{Chadli2014} is also based on \textit{bilinear} matrix inequalities (BMI) while solving the control design problem. Results from \cite{Ji2007} and \cite{Xu} have several disadvantages. In both papers the $\Hinf$ performance analysis for linear discrete-time descriptor systems with uncertainties in all the system's matrices is not represented. In \cite{Xu} the proposed algorithm of $\Hinf$ control problem solution, based on \textit{nonlinear} matrix inequalities approach, is difficult to compute. The approach described in \cite{Ji2007} requires auxiliary matrix variables and fails for high order systems. So the problem of numerically effective robust $\Hinf$ control design for discrete-time systems with norm-bounded uncertainties in terms of linear matrix inequalities (LMI) is an open problem.

This paper presents a new approach to robust $\Hinf$ control problem for discrete-time descriptor systems in terms of LMI. The proposed method is based on the modified BRL, proved in \cite{Feng2013}. The obtained conditions are linear over unknown parameters and numerically effective.

The paper is organized as follows. In the Section \ref{Basics}, basic definitions and background are introduced. Main results of the paper, consisting of the bounded real lemma in terms of LMI and suboptimal robust control procedure for uncertain systems, are represented in the Section~\ref{Mainres}. In the Section, \ref{Hinfex} numerical examples are given.

\section{Preliminaries}
\label{Basics}
In this section, main definitions, concepts, and theorems from the theory of descriptor systems are given~\cite{Dai1989,Xu}.

A state-space representation of discrete-time descriptor systems is
\begin{eqnarray}
\label{Sys1}
  Ex(k+1) &=& A x(k) + B f(k), \\
\label{Sys1_1}
  y(k) &=& C x(k)+D f(k)
\end{eqnarray}
where $x(k) \in \mathbb{R}^n$ is the state, $f(k) \in
\mathbb{R}^m$ and $y(k) \in \mathbb{R}^p$ are the input and output signals, respectively,  $E,\,A,\, B,\, C$ and $D$  are constant real matrices of appropriate dimensions. The matrix $E \in \mathbb{R}^{n \times n}$ is singular, i.e. $\rank(E) = r < n$.

The following denotations are used throughout the paper:
\begin{itemize}
  \item $I_n$ in an identity $n \times n$ matrix;
  \item $\overline{\sigma}(A)$ stands for the maximal singular value of the matrix $A$: $\overline{\sigma}(A)=\sqrt{\rho(A^* A)}$,where $A^*$ is the Hermitian conjugate of the matrix $A$;
  \item $\sym(A)$ stands for symmetrization of the matrix $A$: 
   $\sym(A) = A + A^\rT$.
\end{itemize}

\begin{definition}
The system~\eqref{Sys1} is called admissible if it is
\begin{enumerate}
  \item regular ($\exists\lambda:$ $\det(\lambda E - A) \neq 0$),
  \item causal ($\deg\det(\lambda E-A)=\rank{E}$),
  \item stable ($\rho(E,A)=\max{|\lambda|}_{\lambda\in z |\{\det(z E-A)=0\}}<1$).
\end{enumerate}
\end{definition}

Regularity stands for existence and uniqueness of the solution for consistent initial conditions \cite{Dai1989}. Hereinafter, we suppose that the considered systems are regular.

\begin{definition}
The transfer function of the system~\eqref{Sys1}--\eqref{Sys1_1} is defined by the expression
\begin{equation}\label{sys1tf}
  P(z) = C (z E - A)^{-1} B + D, \mbox{ } z \in \mathbb{C}.
\end{equation}
\end{definition}

For the singular matrix $E$ there exist two nonsingular matrices $\overline{W}$ and $\overline{V}$ such that
$\overline{W}  E \overline{V}  = \diag(I_r,0)$ (see~\cite{Dai1989}).

Consider the following change of variables:
\begin{equation} \label{Vtrans}
    \overline{V} ^{-1}x(k)  = \left[
        \begin{array}{c}
            x_1(k) \\
            x_2(k)
        \end{array}
    \right]
\end{equation}
where $x_1(k) \in \mathbb{R}^r$ and $x_2(k) \in \mathbb{R}^{n-r}$.

By left multiplying of the system~\eqref{Sys1} on the matrix $\overline{W}$ and using the change of variables~\eqref{Vtrans}, one can rewrite the system~\eqref{Sys1}--\eqref{Sys1_1} in the form
\begin{eqnarray}
    \label{ef2}
        x_1(k+1) & = &  A_{11} x_1(k) + A_{12} x_2(k) + B_1 f(k),\\
    \label{ef21}
        0 & = &  A_{21} x_1(k) + A_{22} x_2(k) + B_2 f(k),\\
    \label{ef22}
        y(k) & = & C_1 x_1(k) + C_2 x_2(k) + Df(k)
\end{eqnarray}
where
\begin{multline}
\label{WAV}
    \overline{W}  A \overline{V}   =   \left[
        \begin{array}{cc}
            A_{11} & A_{12} \\
            A_{21} & A_{22}
        \end{array}
    \right],
    \quad
    \overline{W}  B =
    \left[
        \begin{array}{c}
            B_1 \\
            B_2
        \end{array}
    \right],
    \quad \\
    C \overline{V}
    =\left[
        \begin{array}{cc}
            C_1 & C_2
        \end{array}
    \right].
\end{multline}

Matrices $\overline{W}$ and
$\overline{V}$ can be found from the singular value decomposition (SVD)
$$    E
    =
    U
    \diag(S,0)
    H^\rT.
$$
Here $U$ and $H$ are real orthogonal matrices, $S$
is a diagonal $r\times r$-matrix, it is formed by nonzero singular values of the matrix $E$. Then, matrices $\overline{W}$ and $\overline{V}$ may be defined as
$$
      \overline{W}
      =
      \diag(S^{-1}, I_{n-r})
      U^\rT,
      \qquad
      \overline{V}
      =
      H .
$$

The representation~\eqref{ef2}--\eqref{ef22} is called \textit{SVD equivalent form} of the system~\eqref{Sys1}--\eqref{Sys1_1} (see~\cite{Dai1989}).

Consider an input signal in the following form:
\begin{equation}\label{f}
 f(k)=F_c x(k) + h(k)
\end{equation}
where $F_c \in \mathbb{R}^{m\times n}$ is a constant real matrix, $h(k)$ is a new input signal. The equation \eqref{Sys1} turns to
\begin{equation}\label{closesys}
  E x(k+1)=(A+BF_c)x(k)+Bh(k).
\end{equation}
\begin{definition}
The system~\eqref{Sys1} is called causally controllable if there exists a state feedback control in the form~\eqref{f} such that the closed-loop system~\eqref{closesys} is causal.
\end{definition}

The system~\eqref{Sys1} is causally controllable if
$$\rank\left[
         \begin{array}{ccc}
           E & 0 & 0 \\
           A & E & B \\
         \end{array}
       \right]=\rank E+n.
$$

\begin{definition}
The system \eqref{Sys1} is called stabilizable if there exists a state feedback control in the form $f(k)=F_{st} x(k)$ such that the pair $(E, A+BF_{st})$ is stable.
\end{definition}
For more information about causal controllability and stabilizability see~\cite{Dai1989}.

The following results will be used below.

\begin{lemma}\label{Petersen_lemma} (Petersen~\cite{Petersen1987}) \\
Let the matrices $M\in\R^{n\times p}$ and $N\in\R^{q\times n}$ be nonzero, and $G=G^\rT\in\R^{n\times n}$. The inequality
\begin{equation}\label{Petersen_Initial}
  G+M\Delta N+N^\rT \Delta^\rT M^\rT \leqslant 0
\end{equation}
is true for all $\Delta\in\R^{p \times q}$: $\|\Delta\|_2\leqslant 1$ if and only if there exists a scalar value $\varepsilon>0$ such that
\begin{equation}\label{Petersen_Transf}
  G+\varepsilon M M^\rT+ \frac{1}{\varepsilon}N^\rT N \leqslant 0.
\end{equation}
\end{lemma}
\begin{lemma} \label{Schur_lemma}(Schur~\cite{Boyd(1994)}) \\
Let
$$X=\left[
     \begin{array}{cc}
       X_{11} & X_{12} \\
       X_{12}^\rT & X_{22} \\
     \end{array}
   \right]
$$
where $X_{11}$ and $X_{22}$ are square matrices.

If $X_{11}>0$, then $X>0$ if and only if
\begin{equation}\label{Schur}
  X_{22} - X_{12}^\rT X_{11}^{-1}X_{12}>0.
\end{equation}

If $X_{22}>0$, then $X>0$ if and only if
\begin{equation}\label{Schur}
  X_{11} - X_{12} X_{22}^{-1}X_{12}^\rT>0.
\end{equation}
\end{lemma}

The following results are based on the modified bounded real lemma for descriptor systems in SVD equivalent form.

Denote $A_d=\overline{W}  A \overline{V},\,B_d=\overline{W}  B,\,C_d=C \overline{V},\,D_d=D$.

\begin{lemma} \cite{Feng2013} \label{AZDescSysTh1}
The system~\eqref{Sys1} is admissible and $$\| P(z) \|_\infty  < \gamma$$ if there exist matrices $L \in \mathbb{R}^{r\times r}$, $L>0$, $Q\in \mathbb{R}^{r\times r}$, $R \in \mathbb{R}^{r\times (n-r)}$, $S \in \mathbb{R}^{(n-r)\times (n-r)}$, and a sufficiently large scalar $\alpha>0$ such that

\begin{equation}\label{AZLMIDesc}
\left [
\begin{array}{ccccc}
  \Phi_{11} & \Gamma A_d & \Gamma B_d & \Phi_{41}^\rT & 0 \\
  A_d^{\rT} \Gamma^{\rT} & \Phi_{22} & \Pi B_d & A_d^{\rT} \Gamma^{\rT} & \Phi_{52}^\rT \\
  B_d^{\rT} \Gamma^{\rT} & B_d^{\rT} \Pi^{\rT} & -\gamma^2 I_m & B_d^{\rT} \Gamma^{\rT} & \Phi_{53}^\rT \\
  \Phi_{41} & \Gamma A_d & \Gamma B_d & -Q-Q^{\rT} & 0 \\
  0 & \Phi_{52} & \Phi_{53} & 0 & -I_p
\end{array} \right ] <0
\end{equation}

where \\
$\Phi_{11}=-\frac{1}{2}Q-\frac{1}{2}Q^{\rT}$, $\Phi_{22}=\Pi A_d + A_d^{\rT}\Pi^{\rT}-\Theta$, \\ $\Phi_{41}=L-Q-\frac{1}{2}Q^{\rT}$, $\Phi_{52} = C_d+\alpha C_d \Pi A_d$, \\ $\Phi_{53}=D_d + \alpha C_d \Pi B_d$, \\
and
$\Theta=\left[
                \begin{array}{cc}
                  L & 0 \\
                  0 & 0 \\
                \end{array}
              \right],
\,
\Pi=\left[
                \begin{array}{cc}
                0 & 0 \\
                  0 & S \\
                \end{array}
              \right],
\,
\Gamma=\left[
          \begin{array}{cc}
            Q & R \\
          \end{array}
        \right].
$
\end{lemma}

\begin{remark}
In \cite{Feng2013} the modified bounded real lemma is given in assumption that $D = 0$. Using the same algorithm as in~\cite{Feng2013}, one can get the similar conditions for $D \neq 0$, given in the Lemma~\ref{AZDescSysTh1}.
\end{remark}

\begin{remark} \label{alphachoise}
The scalar parameter $\alpha > 0$ in \cite{Feng2013} is assumed to be sufficiently large. This is required to neglect possible risk of conservatism.
\end{remark}

\section{Problem Statement and Main Results}
\label{Mainres}

Consider the following discrete-time descriptor system:
\begin{eqnarray}
\label{Syssynth2}
  E x(k+1) &=& A_\Delta x(k) + B_{\Delta w} w(k)+B_{u} u(k), \\
\label{Syssynth2out}
  y(k) &=& C_{\Delta} x(k)+D_{\Delta w} w(k)
\end{eqnarray}
where $x(k) \in \mathbb{R}^n$ is the state, $w(k) \in
\mathbb{R}^{q}$ is the input disturbance,  $y(k) \in \mathbb{R}^p$ is the output, $u(k) \in \mathbb{R}^{m}$ is the control input.  The matrix $E$ is singular, i.e. $\rank E =r<n$.  $A_{\Delta} = A+M_A\Delta N_A$, $B_{\Delta w} = B_w+M_B\Delta N_B$, $C_{\Delta} = C+M_C\Delta N_C$, $D_{\Delta w } = D_w+M_D\Delta N_D$.

The matrix $\Delta\in\R^{s\times s}$ is unknown norm-bounded, i.e. $\|\Delta\|_2\leqslant 1$ (or Frobenius norm-bounded matrix as $\|\Delta\|_2\leqslant\|\Delta\|_F$). Note that $\|\Delta\|_2:=\overline{\sigma}(\Delta)\leqslant 1$ iff $\Delta^\rT \Delta\leqslant I_s$.

Assume that
\begin{enumerate}
  \item the system~\eqref{Syssynth2} is causally controllable;
  \item the system~\eqref{Syssynth2} is stabilizable;
  \item a scalar value $\gamma > 0$ is known.
\end{enumerate}

The $\Hinf$ performance analysis problem is formulated as follows.

\begin{problem} \label{hinfperfan} Suppose that $B_u = 0$ and the pair $(E,A_\Delta)$ is admissible for all $\Delta$ from the given set. For the given scalar $\gamma>0$ the problem is to check the condition $$\| P_{\Delta}(z) \|_{\infty} < \gamma$$ where
$$P_\Delta(z)=C_\Delta(z E-A_\Delta)^{-1}B_\Delta+D_\Delta.$$
\end{problem}

The $\Hinf$ control design problem is defined as
\begin{problem} \label{hinfcontproblem} For the given scalar $\gamma > 0$ the problem is to find a state-feedback control $$u(k) = F x(k),$$ for which the closed-loop system $$ P_{\Delta}^{cl}(z) = C_\Delta(z E - (A_\Delta + B_u F ))^{-1}B_\Delta+D_\Delta$$ is admissible and $$\| P_{\Delta}^{cl}(z) \|_{\infty} < \gamma.$$
\end{problem}
Introduce the following denotations: \\
 $A_d = \overline{W} A \overline{V}$, $B_{wd} = \overline{W} B_w$, $B_{ud} = \overline{W} B_u$,  $C_d = C \overline{V}$, \\ $D_{wd} = D_w$,  $N_B^d=N_B,$ $M_A^d=\overline{W} M_A,$ \\ $N_A^d=N_A \overline{V},$ $M_C^d= M_C,$ $N_C^d=N_C \overline{V}$ \\
 where $\overline{W}$ and $\overline{V}$ are nonsingular matrices, which transform \eqref{Syssynth2}--\eqref{Syssynth2out} into the SVD equivalent form.

\subsection{Bounded real lemma for descriptor system with norm-bounded parametric uncertainties}
\label{Hinfroban}

First we start from $\Hinf$ robust performance analysis. As mentioned above in Remark \ref{alphachoise} conservative behavior depends on the choice of $\alpha$. However, it's difficult to obtain reliable conditions on $\Hinf$ performance analysis for an uncertain system with a nonzero value of $\alpha$. So, it's assumed that the scalar $\alpha=0$. It means that the conditions obtained below are possibly conservative and not necessary but still applicable. The solution of Problem \ref{hinfperfan} is defined by the following theorem.

%

\begin{theorem} \label{BRLemma_DescSysUnc}
The system~\eqref{Syssynth2}--\eqref{Syssynth2out} is admissible  and $\sn P_\Delta \sn_\infty  < \gamma$ for all $\Delta$ from the given set if there exist a scalar $\varepsilon>0$ and matrices $Q\in \mathbb{R}^{r\times r}$, $R \in \mathbb{R}^{r\times (n-r)}$, $S \in \mathbb{R}^{(n-r)\times (n-r)}$, $L \in \mathbb{R}^{r\times r}$, $L>0$ such that
\begin{equation}\label{LMIUncDesc}
  \left[
    \begin{array}{cc}
      \Sigma+\varepsilon N_1^\rT N_1 & M_1 \\
      M_1^\rT & -\varepsilon I_{4s} \\
    \end{array}
  \right]<0.
\end{equation}
Here
$$\Sigma=\left [
\begin{array}{ccccc}
  \Sigma_{11} & \Gamma A_d & \Gamma B_{wd} & \Sigma_{41}^\rT & 0 \\
  A_d^{\rT} \Gamma^{\rT} & \Sigma_{22} & \Pi B_{wd} & A_d^{\rT} \Gamma^{\rT} & C_d^{\rT} \\
  B_{wd}^{\rT} \Gamma^{\rT} & B_{wd}^{\rT} \Pi^{\rT} & -\gamma^2 I_m & B_{wd}^{\rT} \Gamma^{\rT} & D_d^{\rT} \\
  \Sigma_{41} & \Gamma A_d & \Gamma B_{wd} & -Q-Q^{\rT} & 0 \\
  0 & C_d & D_d & 0 & -I_p
\end{array} \right ],$$
$\Sigma_{11}=-\frac{1}{2}Q-\frac{1}{2}Q^{\rT},$ $\Sigma_{41} =  L-Q-\frac{1}{2}Q^{\rT},$ \\ $\Sigma_{22}=\Pi A_d + A_d^{\rT}\Pi^{\rT}-\Theta,$
$$
M_1=\left[
      \begin{array}{cccc}
        \Gamma M_A^d & \Gamma M_B^d & 0 & 0 \\
        \Pi M_A^d & \Pi M_B^d & 0 & 0 \\
        0 & 0 & 0 & 0 \\
        \Gamma M_A^d & \Gamma M_B^d & 0 & 0 \\
        0 & 0 & M_C^d & M_D \\
      \end{array}
    \right],\,N_1=\left[
                    \begin{array}{ccccc}
                      0 & N_A^d & 0 & 0 & 0 \\
                      0 & 0 & N_B^d & 0 & 0 \\
                      0 & N_C^d & 0 & 0 & 0 \\
                      0 & 0 & N_D & 0 & 0 \\
                    \end{array}
                  \right],
$$
$$\Theta=\left[
                \begin{array}{cc}
                  L & 0 \\
                  0 & 0 \\
                \end{array}
              \right],
\,
\Pi=\left[
                \begin{array}{cc}
                0 & 0 \\
                  0 & S \\
                \end{array}
              \right],
\,\Gamma=\left[
          \begin{array}{cc}
            Q & R \\
          \end{array}
        \right].
$$
\end{theorem}

Transform the expression~\eqref{AZLMIDesc} for the system~\eqref{Syssynth2}--\eqref{Syssynth2out}, noting that $\alpha=0$. We get the inequality

\begin{equation}\label{LMIUncDescProof}
  \Sigma+\sym\left(M_1 \Delta N_1\right)<0.
\end{equation}

Applying Lemmas~\ref{Petersen_lemma} and~\ref{Schur_lemma} to the inequality~\eqref{LMIUncDescProof}, we get
$$\Sigma+\frac{1}{\varepsilon}M_1 M_1^\rT+\varepsilon N_1^\rT N_1<0,$$
$$\Sigma+\varepsilon N_1^\rT N_1-M_1 (-\varepsilon I_{4s})^{-1} M_1^\rT<0,$$
$$\left[
    \begin{array}{cc}
      \Sigma+\varepsilon N_1^\rT N_1 & M_1 \\
      M_1^\rT & -\varepsilon I_{4s} \\
    \end{array}
  \right]<0.
$$
Consequently, the conditions of the Lemma~\ref{AZDescSysTh1} hold true for the system~\eqref{Syssynth2}--\eqref{Syssynth2out}, it means that its norm is bounded by the positive scalar value $\gamma$, i.e. $
\| P_\Delta(z) \|_\infty  < \gamma.
$

\subsection{Robust $\Hinf$ control design procedure}
\label{Hinfrobcont}

The Theorem \ref{BRLemma_DescSysUnc} is based on SVD equivalent form of the system \eqref{Syssynth2}--\eqref{Syssynth2out}. It should be noted, that  feasibility of \eqref{LMIUncDesc} does not depend on choosing $\overline{W}$ and $\overline{V}$, as such matrix transformations are invariant. The derived conditions can be used for solving robust $\Hinf$ control problem~\ref{hinfcontproblem} as follows.

\begin{theorem} \label{syntheorem}
For a given scalar $\gamma > 0$ Problem \ref{hinfcontproblem} is solvable if there exist matrices $L \in \mathbb{R}^{r \times r}$, $L > 0$, $Q \in \mathbb{R}^{r \times r}$, $R \in \mathbb{R}^{r \times (n-r)}$, $S \in \mathbb{R}^{(n-r) \times (n-r)}$, $Z \in \mathbb{R}^{n \times m}$, and a scalar $\epsilon > 0$ such that
\begin{equation}\label{DZsyn1lmi1}
  \left[
    \begin{array}{cc}
      \Lambda +\varepsilon M_2^\rT M_2 & N_2 \\
      N_2^\rT & -\varepsilon I_{4s} \\
    \end{array}
  \right]<0
\end{equation}
where

\begin{eqnarray}
  \label{M2}
  M_2 &=& \left[
                    \begin{array}{ccccc}
                      0 & (M_A^d)^\rT & 0 & 0 & 0 \\
                      0 & 0 & (M_C^d)^\rT & 0 & 0 \\
                      0 & (M_B^d)^\rT & 0 & 0 & 0 \\
                      0 & 0 & (M_D)^\rT & 0 & 0 \\
                    \end{array}
                  \right], \\
  \label{N2}
  N_2 &=& \left[
      \begin{array}{cccc}
        \Gamma (N_A^d)^\rT & \Gamma (N_C^d)^\rT & 0 & 0 \\
        \Pi (N_A^d)^\rT & \Pi (N_C^d)^\rT & 0 & 0 \\
        0 & 0 & 0 & 0 \\
        \Gamma (N_A^d)^\rT & \Gamma (N_C^d)^\rT & 0 & 0 \\
        0 & 0 & (N_B^d)^\rT & (N_D)^\rT \\
      \end{array}
    \right], \\
    \label{DZsyn1lmi2}
  \Lambda &=& \left[
    \begin{array}{ccccc}
      \Lambda_{11} & \Lambda_{21}^\rT & \Lambda_{31}^\rT & \Lambda_{41}^\rT & 0 \\
      \Lambda_{21} & \Lambda_{22} & \Lambda_{32}^\rT & \Lambda_{21} & \Lambda_{52}^\rT \\
      \Lambda_{31} & \Lambda_{32} & -\gamma^2 I_q & \Lambda_{31} & \Lambda_{53}^\rT \\
      \Lambda_{41} & \Lambda_{21}^\rT & \Lambda_{31}^\rT & -(Q + Q^\rT) & 0 \\
      0 & \Lambda_{52} & \Lambda_{53} & 0 & -I_{q} \\
    \end{array}
  \right],
\end{eqnarray}
where \\
$\Lambda_{11} = -\frac{1}{2}Q-\frac{1}{2}Q^\rT,$ $\Lambda_{21} = A_d \Gamma^\rT + B_{ud} Z^\rT \Omega^\rT,$ \\  $\Lambda_{31} = C_d \Gamma^\rT,$  $\Lambda_{41} = L - Q - \frac{1}{2}Q^\rT,$ \\
$\Lambda_{22} = \Pi A_d^\rT + A_d \Pi^\rT + \Phi Z B_{ud}^\rT + B_{ud} Z^\rT \Phi^\rT - \Theta,$   \\ $\Lambda_{32} = C_d \Pi^\rT,$ $\Lambda_{52} = B_{wd}^\rT,$  $\Lambda_{53} = D_{d}^\rT.$ \\
$\Theta = \left[
                     \begin{array}{cc}
                       L & 0 \\
                       0 & 0 \\
                     \end{array}
                   \right],\, \Pi= \left[
                     \begin{array}{cc}
                       0 & 0 \\
                       0 & S \\
                     \end{array}
                   \right],\, \Phi = \left[
                     \begin{array}{cc}
                       0 & 0 \\
                       0 & I_{n-r} \\
                     \end{array}
                   \right],\,$ \\
$\Omega = \left[
                                         \begin{array}{cc}
                                           I_r & 0 \\
                                         \end{array}
                                       \right],\, \Gamma = \left[
                                         \begin{array}{cc}
                                           Q & R \\
                                         \end{array}
                                       \right].$

The gain matrix can be obtained as
\begin{equation}\label{DZfbctrl}
  F = Z^\rT \left[
              \begin{array}{cc}
                Q^{-\rT} & 0 \\
                -S^{-\rT} R^\rT Q^{-\rT} & S^{-\rT} \\
              \end{array}
            \right] \overline{V}^{-1}.
\end{equation}
\end{theorem}

Consider the system, dual to \eqref{Syssynth2}--\eqref{Syssynth2out}. Then, the closed-loop dual system goes to
\begin{eqnarray}
\label{Syssynth3}
  E^\rT x(k+1) &=& (A_\Delta^\rT + F^\rT B_u^\rT) x(k) + C_{\Delta}^\rT w'(k), \\
\label{Syssynth3out}
  y'(k) &=& B_{\Delta w}^\rT x(k)+D_{\Delta w}^\rT w'(k).
\end{eqnarray}

If the system \eqref{Syssynth3}--\eqref{Syssynth3out} is admissible and its $\Hinf$ norm is less than $\gamma$, then Problem \ref{hinfcontproblem} is solved for the system \eqref{Syssynth2}--\eqref{Syssynth2out}. Applying Lemma \ref{BRLemma_DescSysUnc} to the system \eqref{Syssynth3}--\eqref{Syssynth3out} we obtain \eqref{DZsyn1lmi1} where \\
\begin{eqnarray}
\label{Lam21}
  \Lambda_{21} &=& A_d \Gamma^\rT + B_{ud} F_d \Gamma^\rT, \\
  \label{Lam22}
  \Lambda_{22} &=& \Pi A_d^\rT + A_d \Pi^\rT + \Pi F_d^\rT B_{ud}^\rT + B_{ud} F_d \Pi^\rT - \Theta.
\end{eqnarray}

Introduce the following linear change of variables $\left[
                             \begin{array}{cc}
                               Q & R \\
                               0 & S \\
                             \end{array}
                           \right]F_d^\rT=Z
$. It implies that $\left[
                  \begin{array}{cc}
                    Q & R \\
                  \end{array}
                \right]F_d^\rT = \left[
                  \begin{array}{cc}
                    I_r & 0 \\
                  \end{array}
                \right]Z
$ and $\left[
                             \begin{array}{cc}
                               0 & 0 \\
                               0 & S \\
                             \end{array}
                           \right]F_d^\rT = \left[
                             \begin{array}{cc}
                               0 & 0 \\
                               0 & I_{n-r} \\
                             \end{array}
                           \right]Z
$. Substituting it into \eqref{Lam21} and \eqref{Lam22} we get $\Lambda_{21}$ and $\Lambda_{22}$ entries from \eqref{DZsyn1lmi2}.

As the inequality~\eqref{DZsyn1lmi1} holds, then the (1,1) entry implies the matrix $Q$ is invertible. We also suppose, that the matrix $S$ is invertible. If it does not hold, there exists a scalar $\epsilon \in (0,1)$, such that the inequality~\eqref{DZsyn1lmi1} holds true for the scalar matrix $\bar{S} = S + \epsilon I_{n-r}$. So, we can use $\bar{S}$ instead of $S$.

 As pointed out before,  $Q$ and $S$ are invertible. So the feedback gain $F_d$ for the closed-loop system~\eqref{Syssynth3} is $F_d = Z^\rT \left[
                     \begin{array}{cc}
                       Q^{-\rT} & 0 \\
                       -S^{-\rT}R^{\rT}Q^{-\rT} & S^{-\rT} \\
                     \end{array}
                   \right]$. Note that $F_d = F \overline{V}$. By the inverse change of variables we get $F$ from~\eqref{DZfbctrl}. This completes the proof.
%
%
%

\begin{remark}
The conditions of the Theorem \ref{syntheorem} allow to formulate $\gamma$--optimal control problem as follows
$$
\mbox{find: } \min \gamma \mbox{ on the set} \{L, Q, R, S, Z, \epsilon \} \mbox{ subject to \eqref{DZsyn1lmi1}.}
$$
If Problem \ref{hinfcontproblem} is solvable and minimal value of $\gamma=\gamma^{*}$ is found, then $\| P_{\Delta}^{cl}(z) \|_{\infty} \leqslant \sqrt{\gamma^{*}}$.
\end{remark}

\begin{remark} \label{noncoservative}
The obtained solution of robust $\Hinf$ control problem is conservative due to the aforementioned assumption $\alpha=0$. In spite of this fact, in particular cases the solutions with less conservatism can be obtained.

The first case is when the system's parameters satisfy the following conditions:
\begin{eqnarray*}
  \label{ranklim1}
  \rank E^\rT\!\! &=& \rank\left[ E^\rT, C^\rT, N_C^\rT \right], \\
  \label{ranklim2}
  \rank E \mbox{   }&=& \rank\left[ E, B_w, M_B \right].
\end{eqnarray*}

In this case the terms multiplied by $\alpha$ in $\Phi_{52}$ and $\Phi_{53}$ entries in \eqref{AZLMIDesc} are identical to zero for all $\alpha$. In this case the solution of the Problem \ref{hinfcontproblem} is non-conservative.

The second case corresponds to uncertainties only in matrix $A$. Then $\Phi_{52}$ and $\Phi_{53}$ entries for the system in \eqref{AZLMIDesc} become
\begin{eqnarray}
  \Phi_{52} &=& C_{d} + \alpha C_{d} \Pi A_d +  \alpha C_{d} \Pi M_A^d \Delta N_A^d,\\
  \nonumber
  \Phi_{53} &=& D_{wd} + \alpha C_{d} \Pi B_{wd}.
\end{eqnarray}


Matrices $\Lambda$ in \eqref{DZsyn1lmi2}, $M_2$, and $N_2$ from the Theorem \ref{syntheorem} take the following form: \\
$\Lambda_{52} = B_{wd}^\rT + \alpha B_{wd}^\rT \Pi A_d^\rT + \alpha B_{wd}^\rT \Phi^\rT Z B_{ud}^\rT,$ \\
$\Lambda_{53} = D_{d}^\rT + \alpha B_{wd}^\rT \Pi C_d^\rT,$

$M_2=\left[
                    \begin{array}{ccccc}
                      0 & (M_A^d)^\rT & 0 & 0 & 0 \\
                      0 & 0 & 0 & 0 & 0 \\
                      0 & (M_A^d)^\rT & 0 & 0 & 0 \\
                      0 & 0 & 0 & 0 & 0 \\
                    \end{array}
                  \right],$ \\
$N_2=\left[
      \begin{array}{cccc}
        \Gamma (N_A^d)^\rT & 0 & 0 & 0 \\
        \Pi (N_A^d)^\rT & 0 & 0 & 0 \\
        0 & 0 & 0 & 0 \\
        \Gamma (N_A^d)^\rT & 0 & 0 & 0 \\
        0 & 0 & \alpha B_{wd}^\rT \Pi (N_A^d)^\rT & 0 \\
      \end{array}
    \right]. $

Due to the fact that Theorem \ref{syntheorem} provides only sufficient but not necessary conditions, the value $\alpha$ should be chosen carefully.
\end{remark}

\section{Numerical example}
\label{Hinfex}
Consider the following system.
$$
\begin{array}{ccc}
  E = \left[
        \begin{array}{ccc}
          1 & 0 & 0 \\
          0 & 0 & 0 \\
          2 & 0 & 1 \\
        \end{array}
      \right]\!\!\!\!,
   & A = \left[
        \begin{array}{rrr}
          -0.25 & 0 & 0 \\
          -0.5 & 0.5 & 2 \\
          0.75 & -1 & -1.5 \\
        \end{array}
      \right]\!\!\!\!, & B_u = \left[
         \begin{array}{c}
           0 \\
           0 \\
           1 \\
         \end{array}
       \right]\!\!\!\!,
\end{array}
$$
$B_w = \left[
            \begin{array}{rr}
              0 & 0 \\
              0.1 & 0 \\
              0.2 & 0.1 \\
            \end{array}
          \right]\!\!
$, $C = \left[
          \begin{array}{ccc}
            2 & 2 & 0 \\
          \end{array}
        \right]\!\!
$, $D_w = \left[
          \begin{array}{cc}
            0.01 & -0.5 \\
          \end{array}
        \right]\!\!.$

The matrix $A$ is assumed to be uncertain with $\Delta \in [-1; \mbox{ }1]$ and
$$\begin{array}{cc}
    M_A = \left[
         \begin{array}{ccc}
           0.1 & -0.1 & 0.05 \\
         \end{array}
       \right]^\rT\!\!, & N_A = \left[
         \begin{array}{ccc}
           0 & 0.1 & 0.1 \\
         \end{array}
       \right]\!\!.
  \end{array}
$$

The system is causal but not stable. The generalized spectral radius of the nominal system is $\rho(E,A)=2.5$.

The $\Hinf$ control problem is solved using YALMIP. The system has uncertainties only in the matrix $A$. So we can apply algorithm from the Theorem \ref{syntheorem} and from remark \ref{noncoservative}. We are also able to compare the proposed method with methods derived in \cite{Coutinho2014,Ji2007,Xu}

In \cite{Coutinho2014} polytopic-type uncertainties are under consideration. To apply the proposed algorithm the system is transformed to a polytopic-type uncertain system with  $A_1 = \left[
        \begin{array}{rrr}
          -0.25 & -0.1 & -0.01 \\
          -0.5 & 0.51 & 2.01 \\
          0.75 & -1.005 & -1.505 \\
        \end{array}
      \right]$ and $A_2 = \left[
        \begin{array}{rrr}
          -0.25 & 0.1 & 0.01 \\
          -0.5 & 0.49 & 1.99 \\
          0.75 & -0.995 & -1.495 \\
        \end{array}
      \right]$. An auxiliary variable $E_0$ is chosen as $E_0 = \left[
  \begin{array}{ccc}
    0 & e_2 & 0 \\
  \end{array}
\right]^\rT$, where $e_2$ is a scalar. The initialization step for different values of $\mu$ leads to infeasible problem.

As the system has uncertainties only in the matrix $A$, we may apply methods derived in \cite{Ji2007,Xu} to solve robust $\Hinf$ control problem. An auxiliary matrix is chosen similar to the method described above. Algorithm from \cite{Ji2007} leads to infeasible problem. Method from \cite{Xu} is not applicable in YALMIP due to nonlinear constraints.

Minimization of $\gamma$ by the Theorem \ref{syntheorem} method gives us $\gamma_{min}^{(1)}=1.9093$,
$$K_1=\left[
                                    \begin{array}{ccc}
                                      0.2055  &  1.0702  &  1.4786 \\
                                    \end{array}
                                  \right].
$$
The spectral radius of closed-loop systems lies between
$$
0.2473 \leqslant \rho(E,A_{\Delta}+B_u K_1) \leqslant 0.3480.
$$
So, the closed-loop system is admissible for all $\Delta \in [-1; \mbox{ }1]$, and its $\Hinf$ norm is $\|P_{\Delta}^{cl1}(z)\|_{\infty} \leqslant 2.0089$.

Minimization of $\gamma$ using conditions from Remark \ref{noncoservative} for $\alpha=1000$  gives us $\gamma_{min}^{(2)}=1.1848$,
$$K_2=\left[
                                    \begin{array}{ccc}
                                       -0.4887 & 1.8633 & 4.4607 \\
                                    \end{array}
                                  \right].
$$
The spectral radius of closed-loop systems lies between
$$
0.4835 \leqslant \rho(E,A_{\Delta}+B_u K_1) \leqslant 0.5008.
$$
Its $\Hinf$ norm is $\|P_{\Delta}^{cl2}(z)\|_{\infty} \leqslant 1.1044$.

\begin{figure}
  \includegraphics[width=8cm]{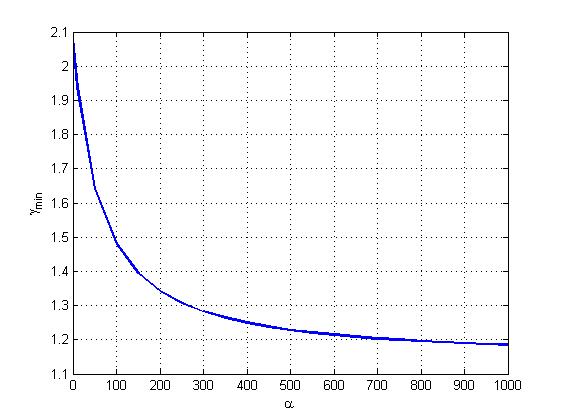}\\
  \caption{Minimized robust $\Hinf$ performance with respect to different $\alpha$.}\label{algamma}
\end{figure}

The evaluation of $\gamma_{min}$ depending on the choice of $\alpha$ is shown on Fig. \ref{algamma}.

\section{Conclusion}\label{conclusion}

In this paper, a new approach to robust $\Hinf$ control problem for discrete-time descriptor systems with norm-bounded uncertainties is presented. The obtained conditions based on strict LMIs are numerically effective. Due to the fact that proposed methods are sufficient but not necessary the conservatism may appear while solving the control design problem. However, the conditions are still applicable. The method of conservatism reduction is proposed for particular case with uncertainties only in the matrix $A$. The effectiveness of proposed methods is illustrated on the numerical example. The numerical example shows that the algorithm is still applicable when other methods fails.

\section*{Acknowledgment}

This work was financially supported by the Government of the Russian Federation (Grant 074-U01) through ITMO Postdoctoral Fellowship program.

\bibliographystyle{plain}        
\bibliography{autosam}

\begin{thebibliography}{99}








\bibitem{Boyd(1994)}
Boyd, S., Ghaoui, L.E., Feron, E., and Balakrishnan, V. (1994). {\it Linear Matrix Inequalities in Systems and Control Theory}, SIAM Studies in Applied Mathematics, Philadelphia, Pennsylvania.


 \bibitem{Chadli2012}
Chadli~M., Darouach~M. (2012). Novel bounded real lemma for discrete-time descriptor systems: Application to $\Hinf$ control design. {\it Automatica, 48},~449--453.


\bibitem{Chadli2014}
Chadli M., Darouach M. (2014). Further Enhancement on Robust $\Hinf$ Control Design for Discrete-Time Singular Systems.
 {\it IEEE Transactions on Automatic Control,~59~(2)},~494--499.



\bibitem{Coutinho2014}
Coutinho D., de Souza C.E., Barbosa K.A. (2014).
Robust $\Hinf$ control of discrete-time descriptor systems in
{\it Proc. of European Control Conference},~1915--1920.


\bibitem{Dai1989}
Dai L. (1989). {\it Singular Control Systems}, Lecture Notes in Control and Information Sciences, New York, Springer-Verlag.

\bibitem{Duan2010}
Duan G.-R. (2010). {\it Analysis and Design of Descriptor Linear Systems},
Advances in Mechanics and Mathematics, Vol. 23, Springer.

\bibitem{Feng2013}
Feng Yu., Yagoubi M. (2013). On state feedback $\Hinf$ control for discrete-time singular systems.
 {\it IEEE Transactions on Automatic Control,~58~(10)},~2674--2679.


\bibitem{Ji2007}
Ji X., Su H., and Chu J. (2007). Robust state feedback $\Hinf$ control for
uncertain linear discrete singular systems. {\it IET Control Theory Appl.,~1~(1)},~195-–200.


\bibitem{Petersen1987}
Petersen, I.R. (1987). A  stabilization  algorithm  for  a  class  of  uncertain linear  systems. {\it Systems and Control  Letters,~8}, ~351--357.

\bibitem{Rehm2002}
 Rehm A., Allg\"{o}wer F. (2002). An LMI Approach towards $\Hinf$ Control of Discrete-time Descriptor Systems in {\it Proc. of American Control Conference},~614--619.





\bibitem{Xu}
Xu S., Lam J. (2006). {\it Robust Control and Filtering of Singular
Systems}, Lecture Notes in Control and Information Sciences,  Berlin, Springer-Verlag.


\bibitem{Yung2008_1}
Yung~C.F.,Wang~C.C., Wu~P.F., Wang~Y.S. (2008).
Bounded real lemma for linear discrete-time descriptor systems in
{\it Proc. 17th IFAC World Congress},~9982--9986.


\bibitem{Yung2008}
 Yung ~C.F. (2008). $\Hinf$ control for linear discrete-time descriptor systems: state feedback and full information cases in {\it Proc. 17th IFAC World Congress},~10003--10008.



\end{thebibliography}

\end{document}